\documentclass{article}
\usepackage[utf8]{inputenc}
\usepackage{geometry}
\usepackage{physics}
\usepackage{dsfont}
\usepackage{amsmath}
\usepackage{amssymb}
\usepackage{graphicx}
\usepackage{babel,blindtext}
\usepackage{xcolor}
\usepackage{hyperref}
\usepackage{cite}
\usepackage{wrapfig}
\usepackage{caption}
\usepackage{subcaption}
\usepackage{bbm}
\usepackage{authblk}
\title{\bf Quantum Correlations and Coherence in a Moving Unruh-deWitt Detector}

\date{}
\author[1]{\small S. Bhuvaneswari}
\author[2]{\small R. Muthuganesan}
\author[3]{\small R. Radha \footnote{Corresponding author}}
\affil[1]{\footnotesize Centre for Nonlinear Science  (CeNSc), PG \& Research Department of Physics, Government College for Women (Autonomous), Kumbakonam, Tamil Nadu, India.}
\affil[2]{\footnotesize Department of Physics, Faculty of Nuclear Sciences and Physical Engineering, Czech Technical University in Prague,
B\u rehov\'a 7, 115 19 Praha 1-Star\'e M\u{e}sto, Czech Republic, Email: rajendramuthu@gmail.com}

\begin{document}

\maketitle
\begin{abstract}
In this paper, we investigate the quantum correlations and coherence of two accelerating Unruh-deWitt detectors coupled to a scalar field in $3+1$ Minkowski space-time. We show that the entanglement is completely destroyed in the limit of infinite acceleration while the local quantum uncertainty and $l_1$-norm of coherence remain nonzero. In addition, we also highlight the role of Unruh temperature and energy spacing of detectors on quantum correlations for different choices of initial states. 
\end{abstract}

\vspace{1cm}
~~~~~~~~~\noindent{\it Keywords}: Quantum Correlation, Coherence, Unruh-deWitt detector, Uncertainty.
\newpage
 \renewcommand{\thefootnote}{*}
\section{Introduction}
Quantum resource theory is an essential tool to understand the nature of physical systems from the perspective of nonlocality and  provides an operational interpretation of various quantum effects \cite{QCbook}. Quantum coherence, superposition, entanglement and quantum correlations beyond entanglement are few notable resources which can prove to be  advantageous over the classical algorithms. Understanding  these resources in  physical systems continue to be a formidable and interesting  task even today. The quantum information processing is usually studied within the framework of   entanglement-versus-separability . The entanglement is considered to be the only early version of nonlocal aspects of  quantum systems \cite{EPR,Schrodinger1,Schrodinger2} and is demonstrated through the violation of Bell inequality \cite{Bell}. The seminal work  of Werner and Almedia et al.  reveals that the entanglement is an incomplete manifestation of nonlocality implying  that quantum correlations cannot only be limited to entanglement \cite{Werner,Almedia} and the separable states can also come in handy in the implementation of some  specific quantum tasks. To study this quantum correlation beyond entanglement, Ollivier and W. H. Zurek introduced a measure called quantum discord which captures the nonlocality of separable (unentangled) states\cite{Ollivier}.

 Other than entanglement, many other types of quantum correlations have been discovered in recent years \cite{Ollivier,Luo2011,GoD}. As a result, a number of quantum correlation concepts have emerged with each being motivated by specific applications in quantum information science competing for recognition as the most accurate quantumness metric. For example, the skew-information and quantum Fisher information play an important role in parameter estimation and bring out  the limitations on the variance of the observable. As stated, each quantum correlation captures different quantumness in quantum systems due to their distinct type of measurement techniques. Among them, skew information based quantum correlation measures such as local quantum uncertainty (LQU) and uncertainty-induced nonlocality (UIN) are   widely used as a quantumness measure to characterize  quantum/physical systems. Girolami et al. \cite{GirolamiPRL2013} have introduced a realizable discord-like nonclassical correlation measure for bipartite systems  known as local quantum uncertainty. Further, this measure quantifies the uncertainty in a quantum state  arising due to its noncommutativity with the measured local observables. The LQU is defined in terms of the skew information. The UIN provides an alternative  to capture the nonlocal effects via local measurements. Both  LQU and UIN possess the closed formula for qubit–qudit ($2\otimes n$ dimensional) systems. Apart from the correlation quantifiers, the quantum coherence is a faithful nonclassicality indicator even for a single qubit system  and is a consequence of superposition  of quantum states. Like  quantum correlations, coherence is also considered to be critical   in the development of quantum technology \cite{Giovannetti,Demkowicz,Lloyd,Huelga,Lambert,Aberg,Lostaglio,Buffoni}. The  quantification of quantum coherence has been recently developed  by Baumgratz et al \cite{Bau}. Recently, the coherence measure based on the $l_1$-norm as  a  valid measure has been introduced and characterized in different contexts \cite{Rana,Chen,Wang}.  Further, the interplay between the quantum coherence measure and correlations is also studied  \cite{Tancoh,Hucoh1}.
 
In relativistic field theory, exploiting the formulation of the detector models, two  well-known effects  such as  Unruh and Hawking effects have been studied. In this formulation, idealized particles are considered as detectors   similar to a two-level system   following the classical worldline and whose internal states  are coupled to the field. In recent times, Unruh-deWitt (UdW) detector model has been devised as a computational element.  The intervention of the environment on the system may cause decoherence and degrades the unique features of quantum system. Assuming that the quantum fluctuations of the background field plays the role of an environment, the Unruh temperature $T_U = a/2\pi $($a$ is acceleration) plays the role of decoherence and destroys the properties of coupled UdW detectors. Here,  the pair of detectors is considered as an open system and its dynamics  is governed by the master equation \cite{Op1}. Recently, different approaches have been employed to study the quantum effects like entanglement, quantum correlations and coherence  \cite{Entg6,Entg8,Entg9,Huang1,Huang2,Huang3}.  Moreover, the quantum correlation and entropic uncertainty relation are also investigated in a multipartite scenario\cite{multiparty1,multiparty2}. 

In this article, we focus on the generation and retention of quantumness in two-UdW detectors interacting with the scalar field. To quantify the quantumness of the detectors, we use the entanglement, local quantum uncertainty, uncertainty-induced nonlocality and $l_1$-norm of coherence.  We show that the entanglement decreases monotonically and vanishes at a finite temperature. The coherence and LQU exhibit a remarkable difference namely, revival of quantum correlation after vanishing at a finite time. Further, we study the dynamics of a product and an incoherent state and the investigation reveals that the dynamics generates the  correlation and coherence from the product and incoherent state respectively.

The  paper is structured  as follows. In Sec. \ref{corr}, we review the quantifier of quantum correlation and coherence measures  . In Sec.  \ref{Sec3}, we introduce the physical model under our consideration and thermal state of two-UdW detectors. The investigations on quantum correlation and coherence are presented in Sec.  \ref{Res}. Finally, the  conclusions are drawn in Sec.  \ref{conc}.
 
\section{Quantum Correlation Measures}
\label{corr}

In this section, we review some of the popular quantum correlation measures to be investigated in this article. For this purpose, we consider an arbitrary bipartite state $\rho$ shared by the subsystems $a$ and $b$ in the separable Hilbert space $\mathcal{H}^a \otimes \mathcal{H}^b$.

\subsection{Entanglement}
Entanglement  is a nonnegative real function of a state $\rho$ which cannot increase under the local operations and classical communication (LOCC) and is zero for unentangled states. To quantify the amount of entanglement associated with the  two-qubit physical system $\rho$ under consideration, various measures have  been introduced. The concurrence is the most popular measure of entanglement for mixed bipartite systems and is defined as \cite{Hill}
\begin{align}
C(\rho)= \text{max}\{0,~ \lambda_1-\lambda_2-\lambda_3-\lambda_4\},
\end{align}
where $\lambda_i$ are eigenvalues of the matrix $R=\sqrt{\sqrt{\rho}\tilde{\rho}\sqrt{\rho} }$  $(\lambda_1\geq \lambda_2\geq \lambda_3\geq \lambda_4)$. Here, $\tilde{\rho}=(\sigma_y \otimes \sigma_y) \rho^* (\sigma_y \otimes \sigma_y)$ is a spin flipped matrix with * denoting the complex conjugate in the computational basis. The function $C(\rho)$ ranges from 0 to 1 and its minimal and maximal values correspond to separable and entangled states respectively.

 \subsection{Local quantum uncertainty}

The local quantum uncertainty (LQU) is a more reliable measure of the quantumness of bipartite states which goes beyond entanglement. In recent times, researchers have paid wide attention to this discord-like measure. This is essentially due to its  easy computation and the fact that it enjoys all necessary properties of being a faithful measure of quantum correlation. It is shown that LQU is non-zero for the separable state even in the absence of entanglement. For a bipartite state $\rho$, the LQU is defined as the minimal skew information attainable with a single local measurement. Mathematically, it is defined as \cite{GirolamiPRL2013},
\begin{align}
\mathcal{Q}_{\mathcal{I}}(\rho)=~^{\text{min}}_{H^{a}} ~~\mathcal{I}(\rho, H^a\otimes\mathds{1}^b),
\end{align}
where $H^a$ is anylocal observable on the subsystem $a$, $\mathds{1}^b$ is the $2\times 2$ identity operator acting on the system $b$ and
\begin{align}
\mathcal{I}(\rho)=-\frac{1}{2}\text{Tr}\left( [\sqrt{\rho}, H^a\otimes\mathds{1}^b]^2 \right)
\end{align}
is the skew information which provides an analytical tool to quantify the information content in the state $\rho$ with respect to the observable $H^a$, $[\cdot, \cdot]$ is the commutator operator.  Here, the  information  content  of $\rho$ about $H^a$ is  quantified  by  how  much  the  measurement  of $H^a$ on  the  state  is  uncertain. The measurement outcome is certain if only if the state is an eigenvector of $H^a$. On the other hand, if  it  is  a  mixture of eigenvectors of $H^a$,  the uncertainty is only due to the imperfect knowledge of the state. For pure bipartite states, the local quantum uncertainty reduces to the linear entropy of entanglement and vanishes for classically correlated states. For any $2\times n$ dimensional bipartite system, the closed formula of LQU is computed as
\begin{align}
\mathcal{U}(\rho)=~ 1- \text{max}\{\omega_1,\omega_2,\omega_3 \}.
\end{align}
Here, $\omega_i$ are the eigenvalues of matrix $W$ and the matrix elements are defined as
\begin{align}
\omega_{ij}=\text{Tr}[\sqrt{\rho}(\sigma_i^a\otimes\mathds{1}^b)\sqrt{\rho}(\sigma_j^a\otimes\mathds{1}^b)],~~~~~\text{with}~~~ i,j=1,2,3,
\end{align}
where $\sigma_i$ represents the Pauli spin matrices.

\subsection{Uncertainty-induced nonlocality}

Next, we employ another important skew information based measure ,namely  uncertainty-induced nonlocality (UIN) and it can be considered as an updated version of measurement-induced nonlocality (MIN) \cite{Luo2011}. It is defined as \cite{UIN}
\begin{align}
\mathcal{Q}(\rho)=~^{\text{max}}_{H^{a}} ~~\mathcal{I}(\rho, H^a\otimes\mathds{1}^b).
\end{align}
This measure also satisfies all the necessary axioms of a valid measure of bipartite quantum correlation and is reduced to entanglement monotone for $2\times n$ dimensional pure states.
It also possesses a closed formula \cite{UIN}.
\subsection{$C_{l_1}$-norm coherence}
Quantum coherence is an important resource for information processing and a manifestation of the quantum superposition principle. Recently, its quantification has been formulated and  a set of conditions to be satisfied by any proper measure of coherence has been identified \cite{Rana,Chen,Wang}.  The distance based quantum coherence measure
\begin{align}
    C(\rho)=~^{\text{min}}_{\delta\in \mathcal{I}} ~d(\rho, \delta)
\end{align}
is the minimal distance between $\rho$ and a set of incoherent quantum states $\delta\in \mathcal{I}$. 
 Recently, a faithful measure of coherence  using $l_1$-norm has been identified  and is defined as the sum of absolute values of all off-diagonal elements of $\rho$ \cite{Bau}
\begin{align}
C_{{l_1}}(\rho)=\sum_{i\neq j}|\rho_{ij}|.
\end{align}
The above definition is a basis dependent measure and is crucial in the identification of phase transition in physical models \cite{LPT1,LPT2,LPT3}.


\section{Physical Model}\label{Sec3}
To understand the behaviors of quantum correlations in two accelerating UdW detectors in $3+1$-dimensional Minkowski space-time \cite{Entg6}, we first introduce the Hamiltonian of the  physical system under  consideration. Considering the detector as a two-level atomic system and a valid qubit system, we consider the two accelerating UdW detectors in $3+1$-dimensional Minkowski space time as an open quantum system for our investigation. The total Hamiltonian of the combined system becomes 
\begin{align}
H=\frac{\omega}{2}\Sigma_3+H_\Phi+\mu H_I             \label{eq1}
\end{align}
where  $\omega$  is the energy level spacing of the atom and $\Sigma_3$ is one of the symmetrized bipartite operators. $\Sigma_i\equiv\sigma_i^a\otimes\mathbf{1}^b+\mathbf{1}^a\otimes\sigma_i^b$ is defined by Pauli matrices $\{\sigma^{(\alpha)}_i|i=1,2,3\}$ with superscripts $\{\alpha=a, b\}$ labelling distinct atoms. $H_\Phi$ is the Hamiltonian of free massless scalar fields $\Phi(t,\mathbf{x})$ satisfying standard Klein-Gordon relativistic equation. The interaction Hamiltonian between the atoms and the fluctuating field bath in a dipole form  can be written as  $H_I= (\sigma_2^{(a)}\otimes\mathbf{1}^{(b)})\Phi(t,\mathbf{ x}_1)+(\mathbf{1}^{(a)}\otimes\sigma_2^{(b)})\Phi(t,\mathbf{ x}_2)$ \cite{Op2}.

Here, we consider the coupling between the detectors and the environment ($\mu \leq 1$) and the initial state of the composite system  is $\rho_{tot}(0)=\rho_{ab}(0)\otimes |0\rangle \langle 0|$, with $\rho_{ab}(0)$ being the initial state of the detectors and $|0\rangle$ is the field vacuum (environment). The time evolution of $\rho_{tot}(0)$ is unitary  governed by von Neumann equation $\dot{\rho}_{tot}(\tau)=-i[H,\rho_{tot}(\tau)]$, where $\tau$ is the proper time of the atom. Due to the environment decoherence or dissipation on the system $\rho_{ab}$, the density matrix  is governed by a Lindblad master equation and evolves non-unitarily in the following form \cite{Op3,Op4}
\begin{align}
\frac{\partial\rho_{ab}(t)}{\partial t}=-i[H_{\tiny\mbox{eff}},\rho_{ab}(t) ]+\mathcal{L}\left[\rho_{ab}(t)\right] 
\label{eq2}
\end{align}
where
\begin{align}
\mathcal{L}\left[\rho\right]=\sum_{\substack{i,j=1,2,3\\   \alpha,\;\beta=a,b}}\frac{C_{ij}}{2}\big[2\sigma_j^{(\beta)}\rho_{ab}\sigma_i^{(\alpha)}-\{\sigma_i^{(\alpha)}\sigma_j^{(\beta)},\rho_{ab}\}\big]       \label{eq3}
\end{align}
describes the evolution due to the interaction between the detectors and external field. 

After introducing the Wightman function of scalar field $G^{+}\left(x, x^{\prime}\right)=\left\langle 0\left|\Phi(x) \Phi\left({x}^{\prime}\right)\right| 0\right\rangle$, its Fourier transform
\begin{align}
\mathcal{G}(\lambda)=\int_{-\infty}^{\infty}d\tau~e^{i\lambda\tau}G^+(\tau)= \int_{-\infty}^{\infty}d\tau~e^{i\lambda\tau}\left\langle\Phi(\tau)\Phi(0)\right\rangle              
\label{eq4}
\end{align}
determines the coefficients $C_{ij}$ by a decomposition
\begin{align}
C_{ij}=\frac{\gamma_+}{2}\delta_{ij}-i\frac{\gamma_-}{2}\epsilon_{ijk} \delta_{3,k}+\gamma_0\delta_{3,i}\delta_{3,j}          \label{eq5}
\end{align}
where
\begin{align}
\gamma_\pm= \mathcal{G}(\omega)\pm \mathcal{G}(-\omega),~~~\gamma_0=\mathcal{G}(0)-\gamma_+/2.         \label{eq6}
\end{align}
Moreover, the interaction with external scalar field would also induce a Lamb shift contribution for the effective Hamiltonian of the detector $H_{\mbox{\tiny eff}}=\frac{1}{2}\tilde{\omega}\sigma_3$ in terms of a renormalized frequency $\tilde{\omega}=\omega+i[\mathcal{K}(-\omega)-\mathcal{K}(\omega)]$ where $\mathcal{K}(\lambda)=\frac{1}{i\pi}\mbox{P}\int_{-\infty}^{\infty}d\omega\frac{\mathcal{G}(\omega)}{\omega-\lambda}$ is a Hilbert transform of Wightman function. 

Following a trajectory of the accelerating detectors, one can find that the field Wightman function fulfills the Kubo-Martin-Schwinger (KMS) condition, i.e., $G^{+}(\tau)=G^{+}(\tau+i \beta)$, where $\beta\equiv1/T_U=2\pi/a$. Translating it into frequency space, one finds that
\begin{align}
\mathcal{G}(\lambda)=e^{\beta\omega}\mathcal{G}(-\lambda).        \label{eq7}
\end{align}
Using translation invariance $\langle 0|\Phi(x(0)) \Phi(x(\tau))| 0\rangle=\langle 0|\Phi(x(-\tau)) \Phi(x(0))| 0\rangle$ and  after some algebraic manipulations, we find that eq.(\ref{eq6}) can be resolved as

\begin{align}
\gamma_+=&\int_{-\infty}^{\infty}d\tau~e^{i\lambda\tau}\langle 0|\left\{\Phi(\tau),\Phi(0)\right\}|0\rangle =\left(1+ e^{-\beta\omega}\right) \mathcal{G}(\omega), ~~~~~~~~\\
\gamma_-=&\int_{-\infty}^{\infty}d\tau~e^{i\lambda\tau}\langle 0|\left[\Phi(\tau),\Phi(0)\right]|0\rangle =\left(1- e^{-\beta\omega}\right) \mathcal{G}(\omega).
\label{eq8}   
\end{align}
It should be  be noted that eq. (\ref{eq8}) holds true for generic interacting fields. Considering the two-atom state in Bloch representation, we compute t he final equilibrium state  of two-UdW detectors asymptotically as 
\begin{align}
\rho_{ab}=\left(\begin{array}{cccc}
\varrho_{11} & 0 & 0 & 0 \\
0 & \varrho_{22} & \varrho_{23} & 0 \\
0 & \varrho_{23} & \varrho_{22} & 0 \\
0 & 0 & 0 & \varrho_{44} 
\end{array}\right)\label{eq11}
\end{align}
where the matrix elements are 
\begin{eqnarray}
  \displaystyle \varrho_{11} &=&\frac{(3+\Delta_0)(\gamma-1)^{2}}{4\left(3+\gamma^{2}\right)},~~~~~~~~
\displaystyle \varrho_{44}=\frac{(3+\Delta_0)(\gamma+1)^{2}}{4\left(3+\gamma^{2}\right)}, \nonumber\\
\displaystyle \varrho_{22}&=& \frac{3-\Delta_0-(\Delta_0+1) \gamma^{2}}{4\left(3+\gamma^{2}\right)},~~~~
\displaystyle \varrho_{23}= \frac{\Delta_0-\gamma^{2}}{2\left(3+\gamma^{2}\right)},    \label{eq12}
\end{eqnarray}
with the parameter 
\begin{align}
\gamma\equiv\gamma_-/\gamma_+=\frac{1- e^{-\beta\omega}}{1+ e^{-\beta\omega}}=\tanh(\beta\omega / 2).      \label{eq9}
\end{align}
It should be mentioned that the parameter depends solely on the Unruh temperature $T_U$ and characterizes the thermal nature of Unruh effect. Further, the dimensionless parameter $\Delta_0=\sum_i\text{Tr}[\rho_{ab}(0)\sigma^{a}_i\otimes\sigma^{b}_i]$ provides the choice of initial state and ranges as $-3\leqslant\Delta_0\leqslant1$.
\begin{figure}[t]
\begin{center}
\includegraphics[width=.45\textwidth]{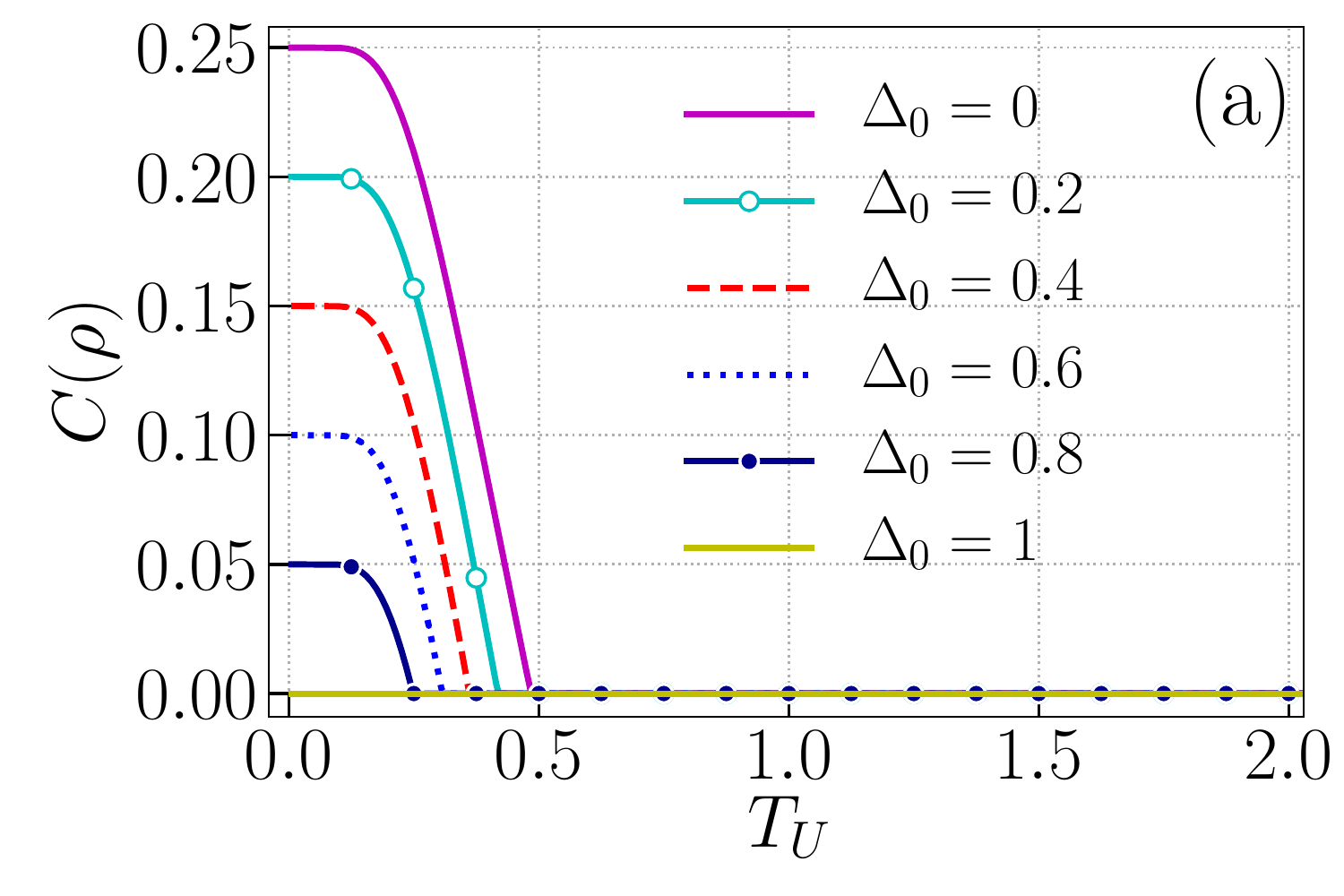}
\includegraphics[width=.45\textwidth]{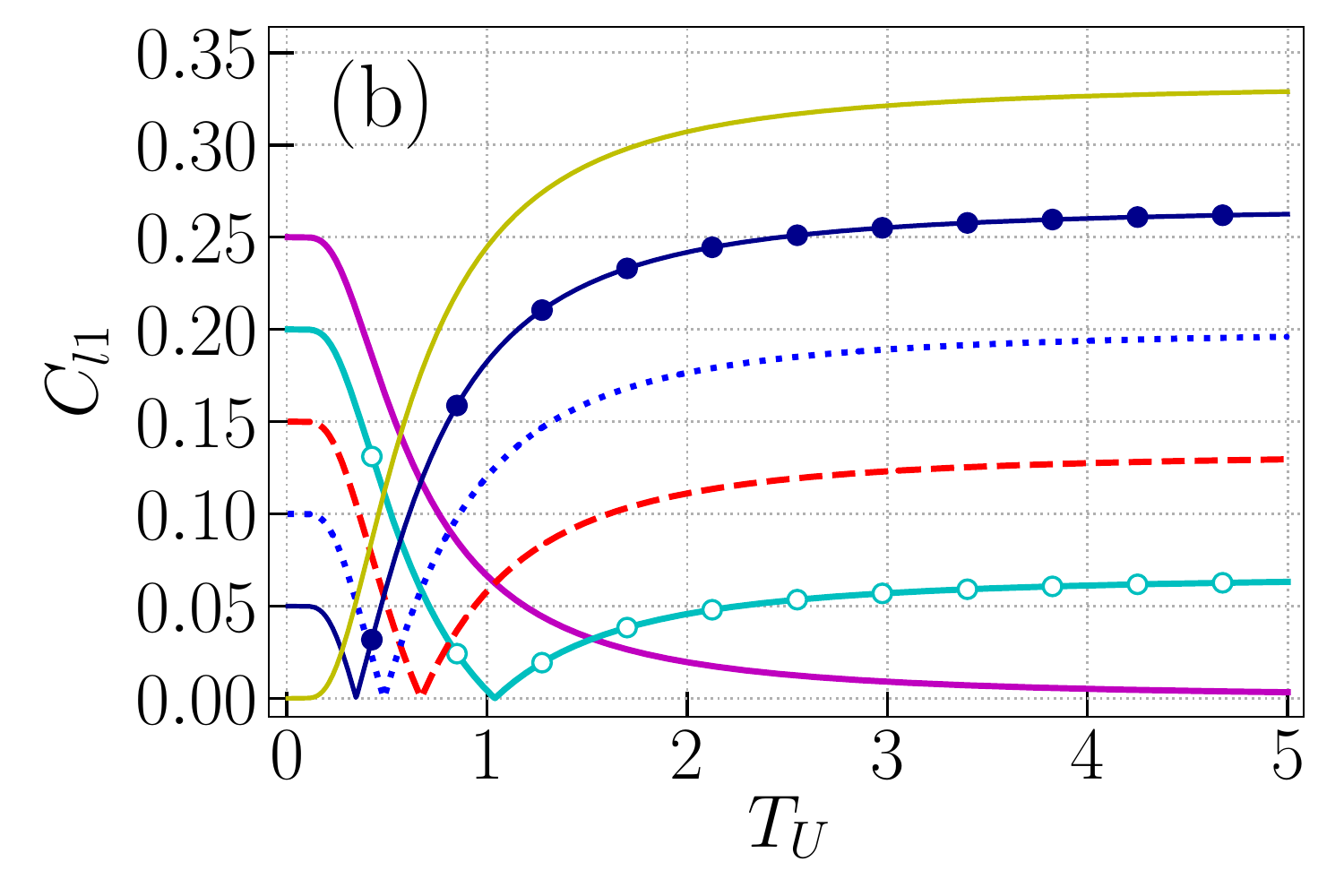}
\includegraphics[width=.45\textwidth]{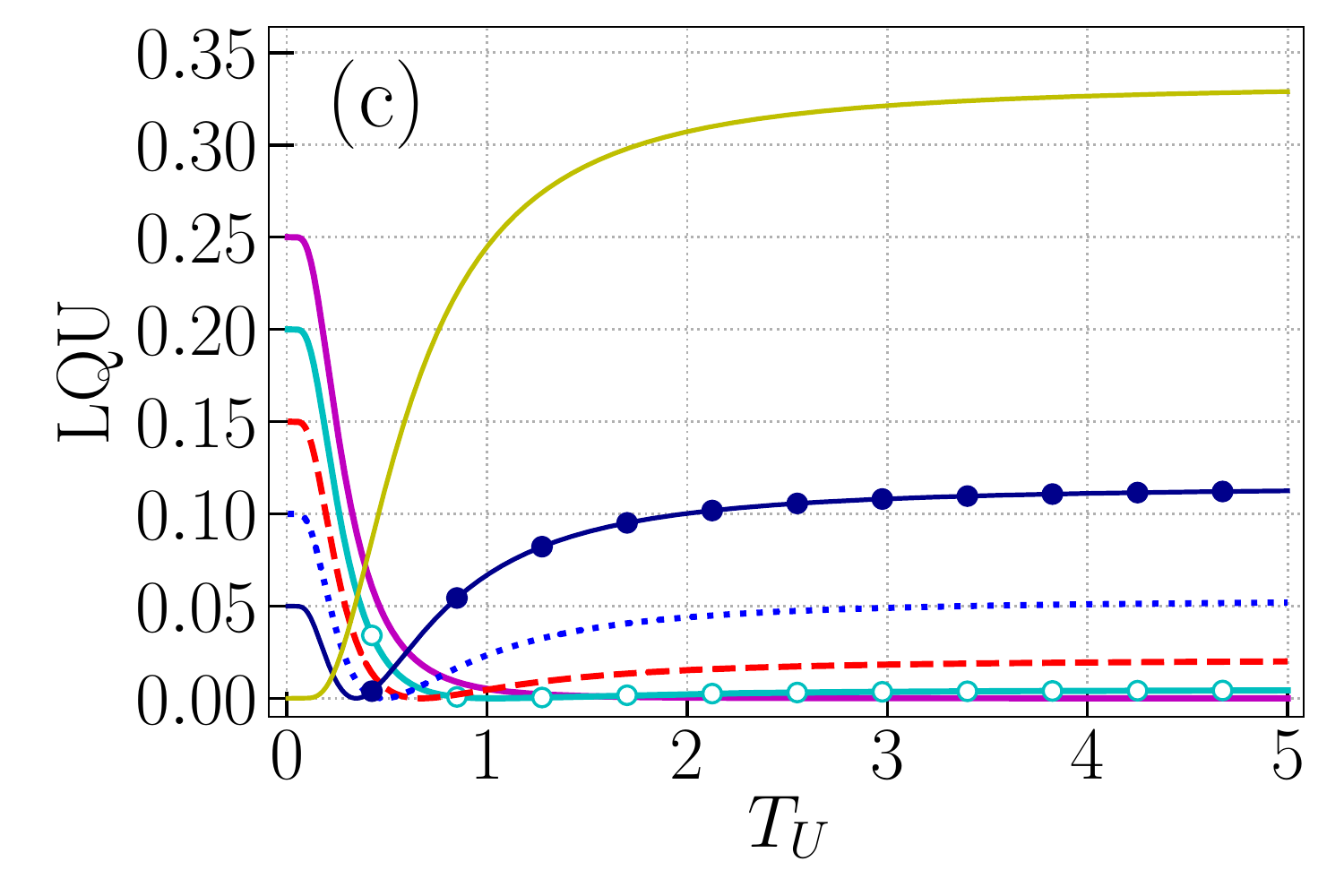}
\includegraphics[width=.45\textwidth]{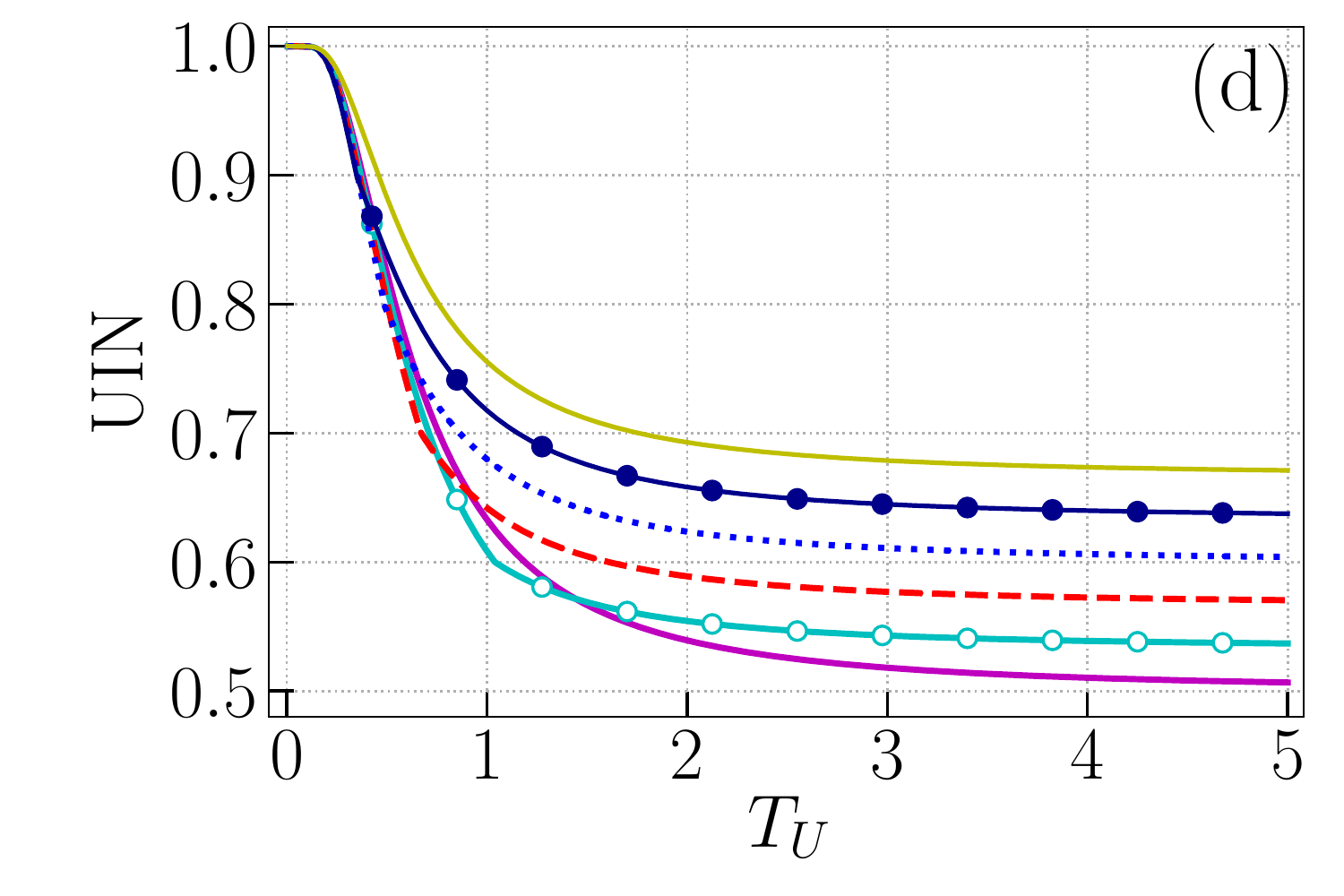}
\caption{Thermal quantum correlation quantified by (a) Entanglement, (b) $l_1$ norm of coherence, (c) LQU and (d) UIN of UdW detector as a function of Unruh temperature $T_U$ for different initial states for  $\omega=1$.}
\label{fig1}
\end{center}
\end{figure}


\section{Results and Discussions}
\label{Res}
\begin{figure}[t]
\begin{center}
\includegraphics[width=.45\textwidth]{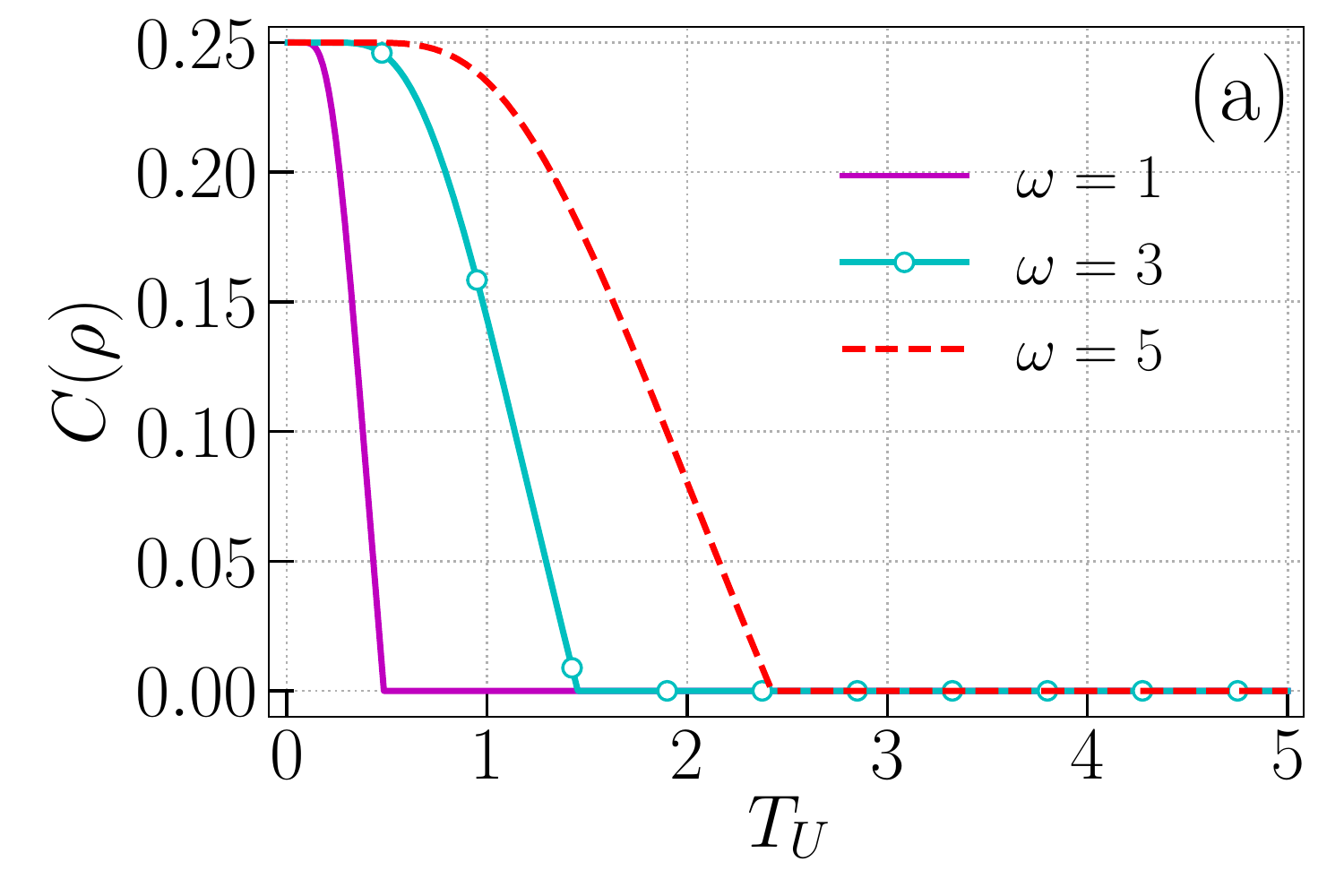}
\includegraphics[width=.45\textwidth]{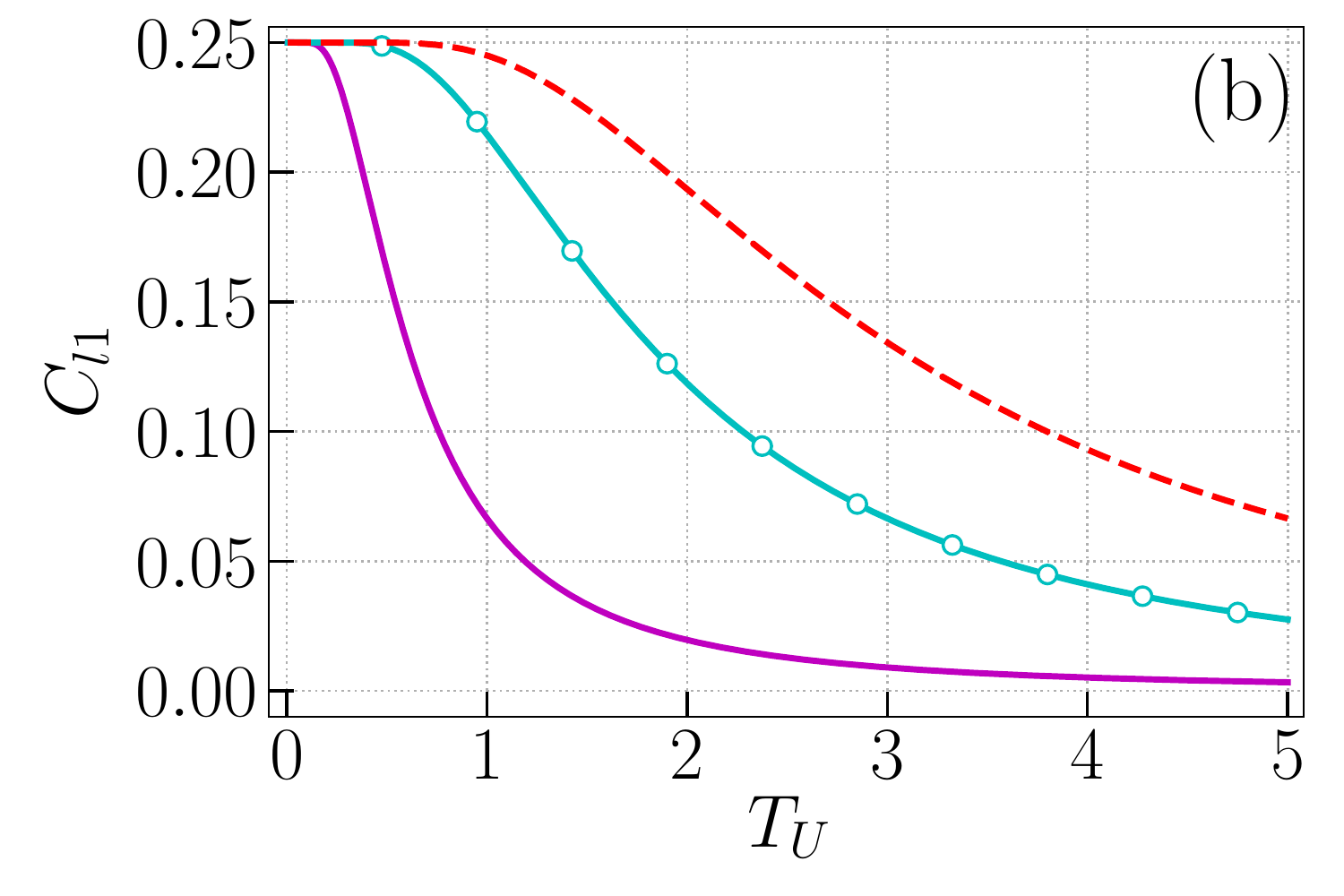}
\includegraphics[width=.45\textwidth]{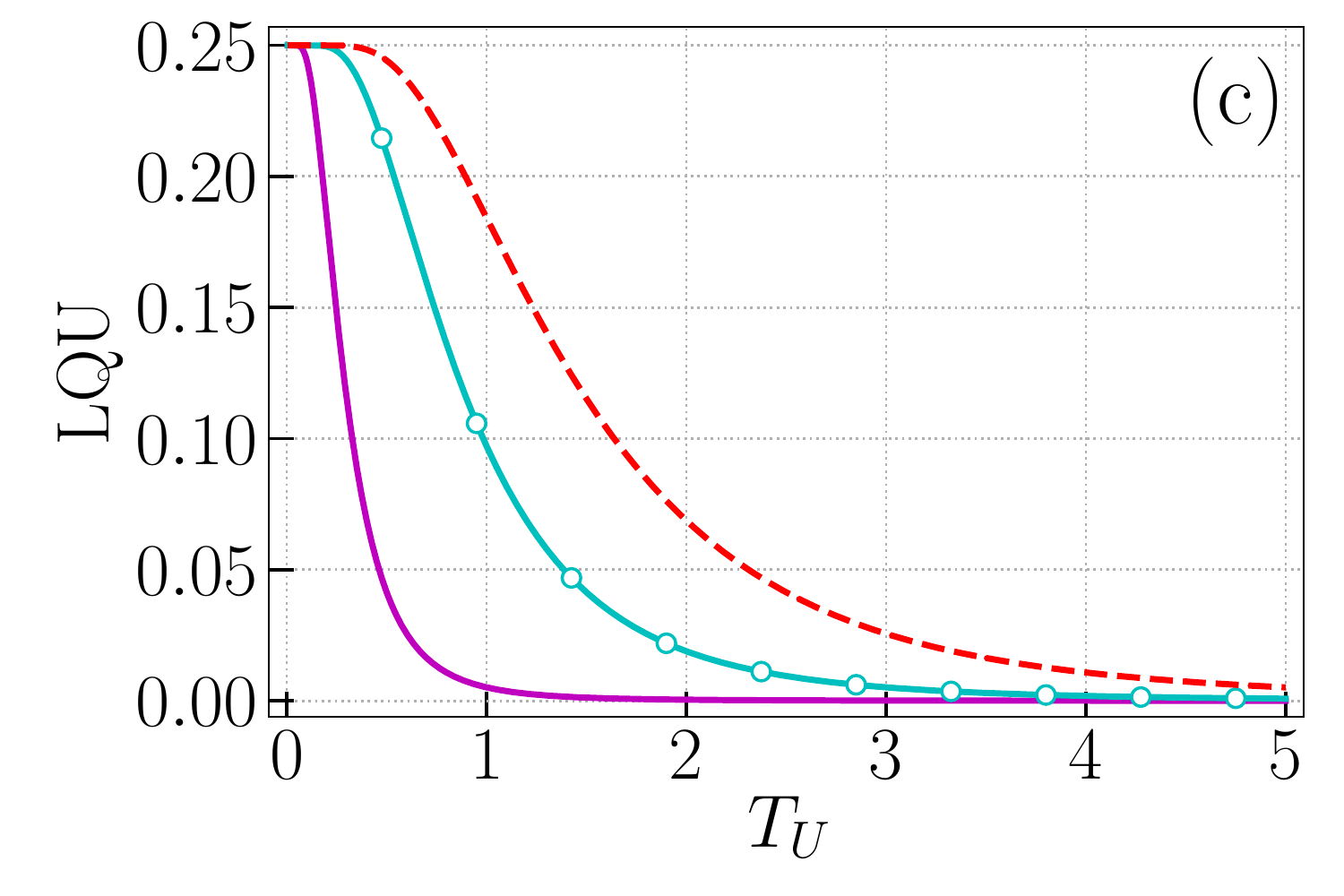}
\includegraphics[width=.45\textwidth]{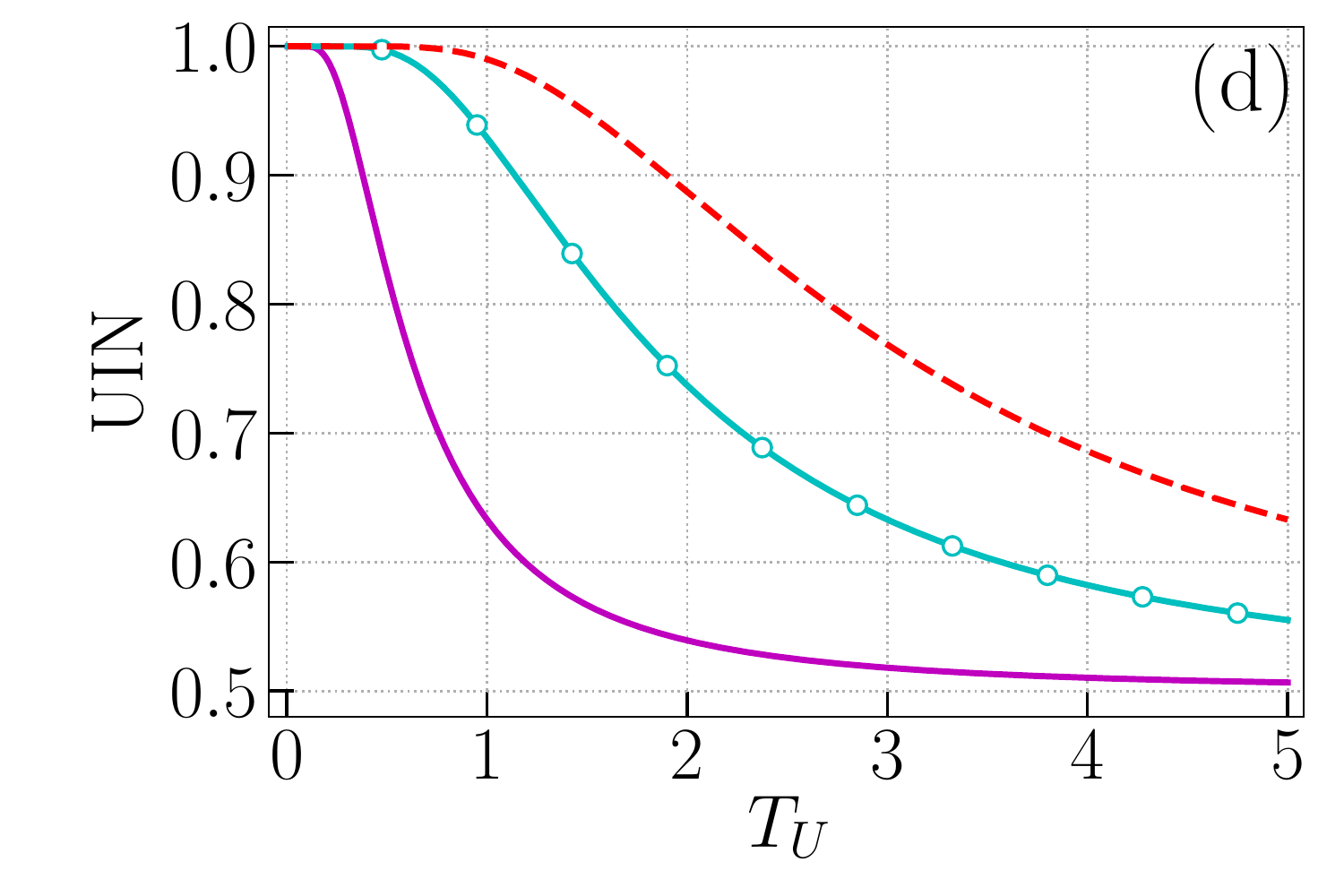}
\caption{Behaviors of quantum correlation measures (a) Entanglement (b) $l_1$ norm of coherence, (c) LQU and (d) UIN of UdW detector as a function of Unruh temperature $T_U$ for $\Delta_0=0$.}

\label{fig2}
\end{center}
\end{figure}

\begin{figure}[t]
\begin{center}
\includegraphics[width=.45\textwidth]{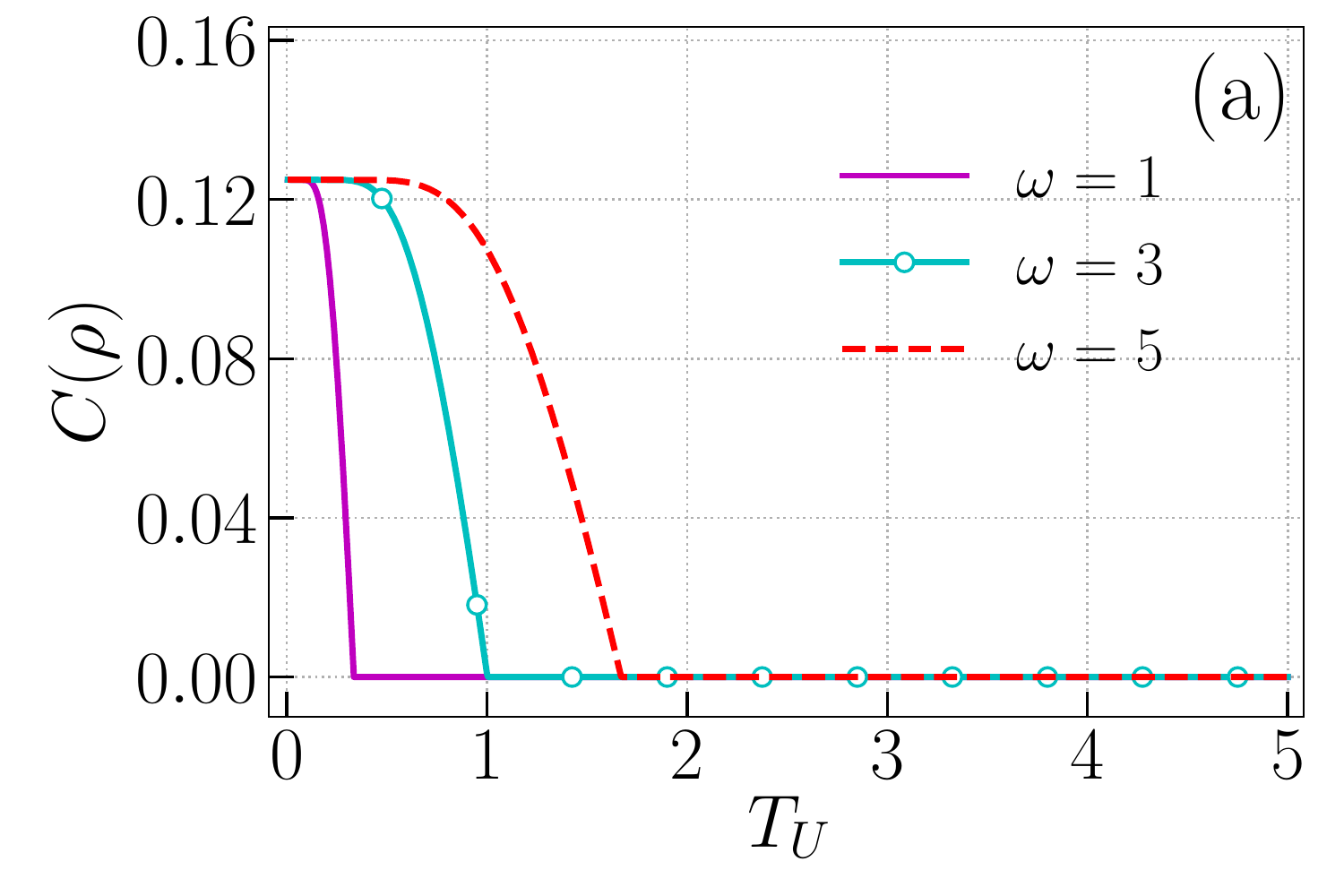}
\includegraphics[width=.45\textwidth]{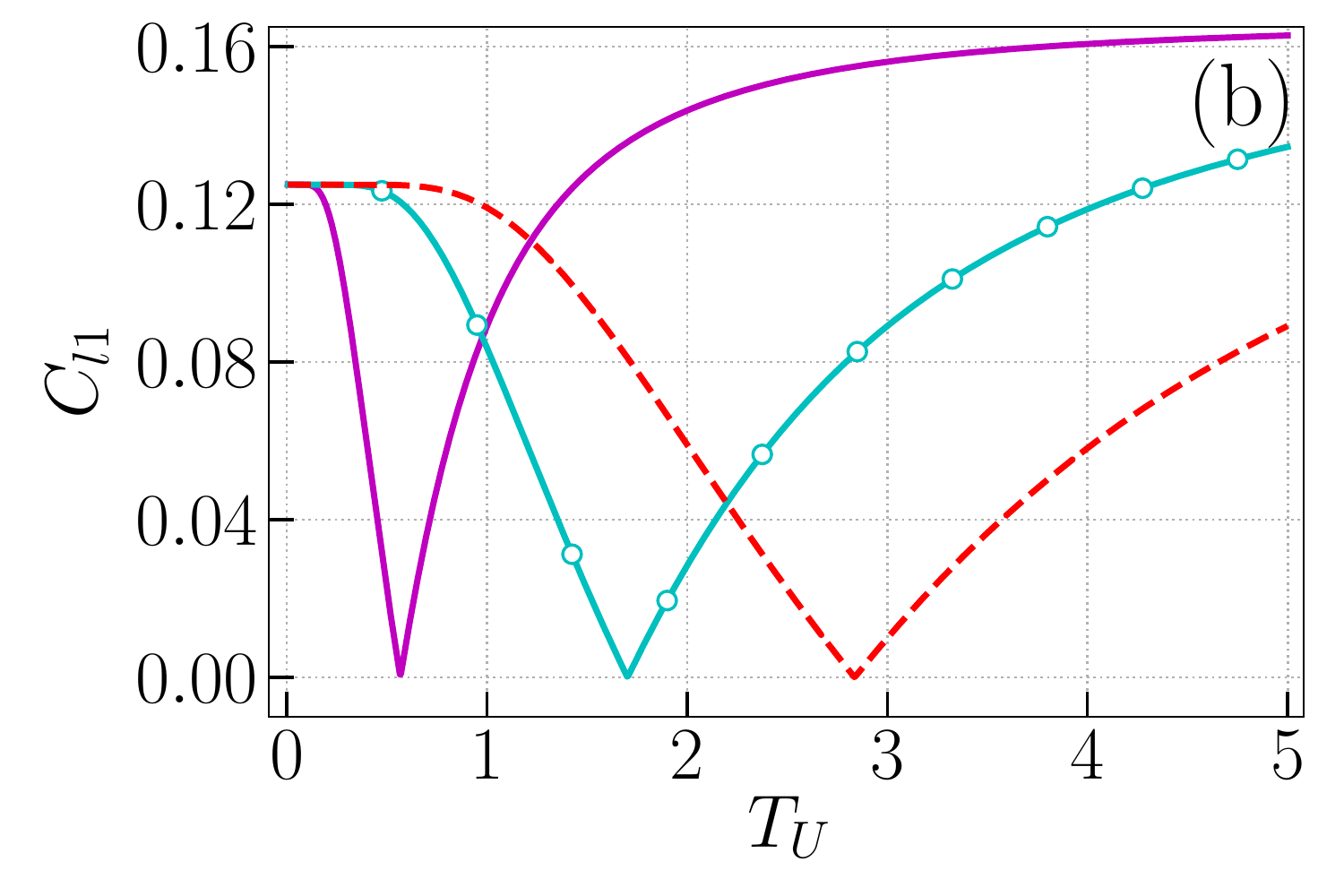}
\includegraphics[width=.45\textwidth]{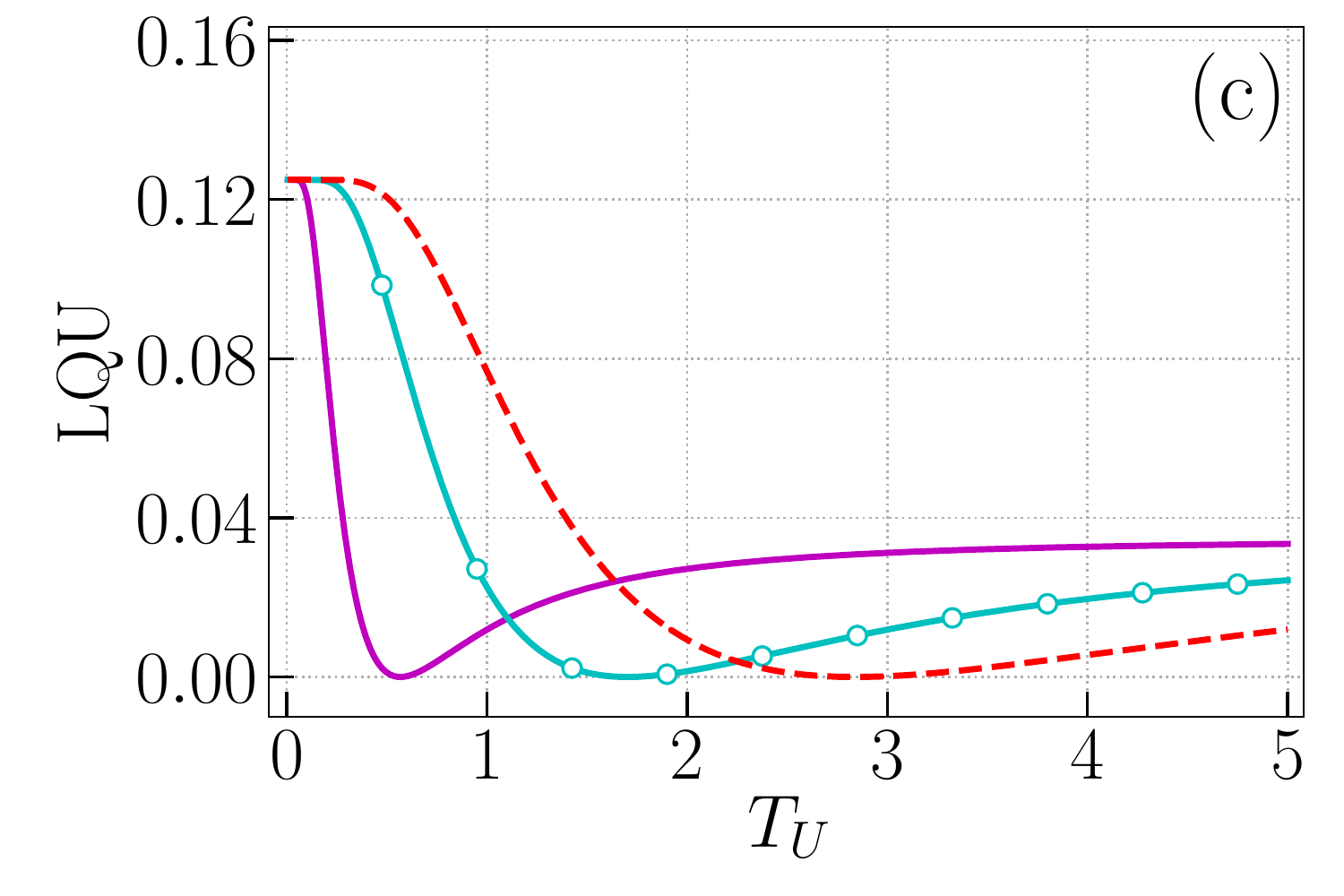}
\includegraphics[width=.45\textwidth]{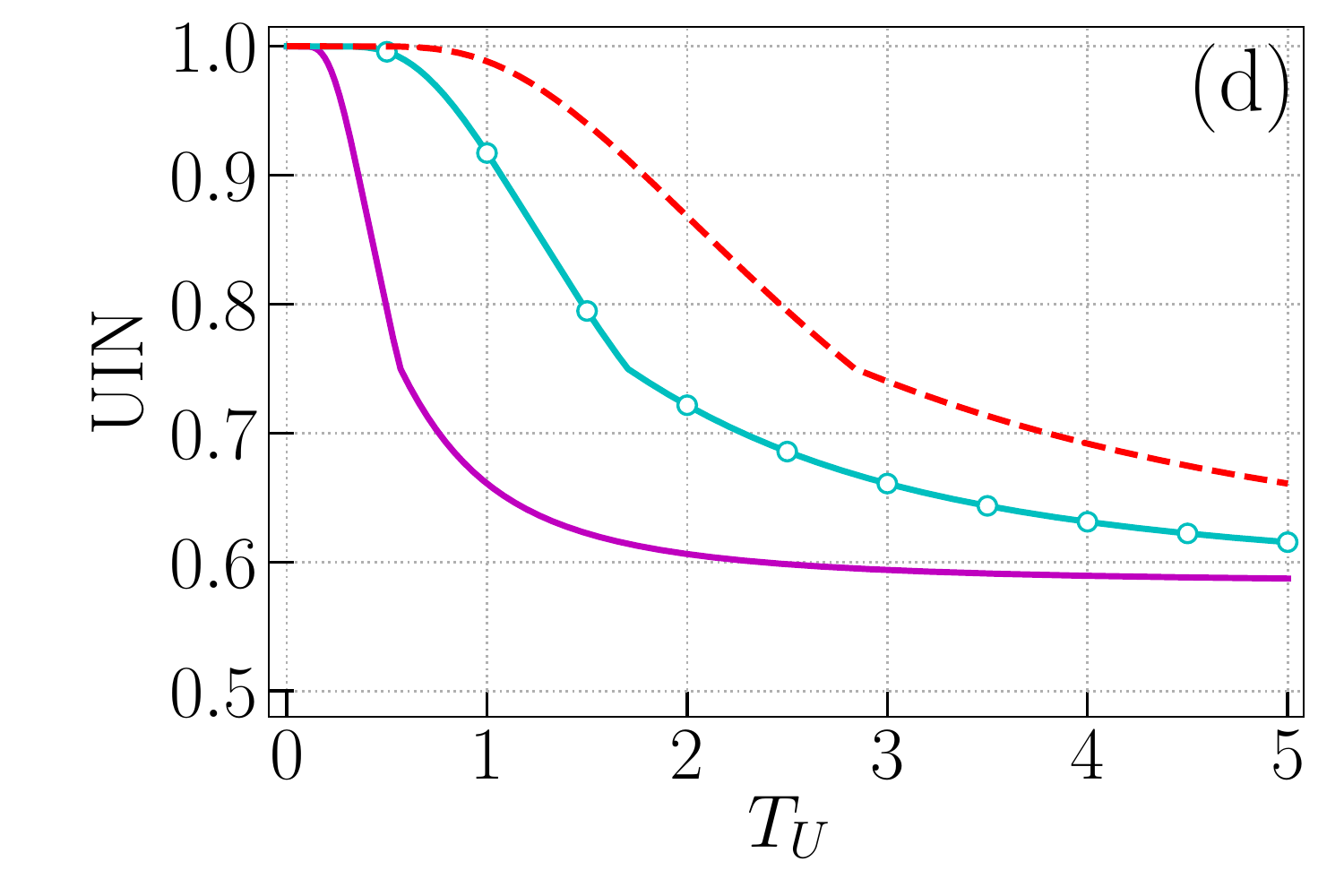}
\caption{Behaviors of quantum correlation measures (a) Entanglement (b) $l_1$ norm of coherence, (c) LQU and (d) UIN of UDW detector as a function of Unruh temperature $T_U$ for different initial states for  $\Delta_0=0.5$.}
\label{fig3}
\end{center}
\end{figure}

We now study the dynamics of quantum correlations measures quantified in terms of entanglement, $C_{l_1}$-norm of coherence , local quantum uncertainty and uncertainity induced nonlocality. 

In general, Unruh effect is recognized as an environment decoherence. Hence, we study the quantum correlations as a function of Unruh temperature $T_U$. In Fig. (1), we plot the quantum correlation measures as a function of Unruh temperature for different choices of initial conditions $\Delta_0$ with a fixed value of $\omega$. For a given initial state, the entanglement is maximum at $T=0$. As temperature increases, the concurrence is a monotonically decreasing function of Unruh temperature $T_U$ and abruptly vanishes . The entanglement completely vanishes for $\Delta_0=1$.  Further, one can observe that the increase of  $\Delta_0$ also reduces the quantum of entanglement between the detectors. In order to compare the entanglement with other correlation measures such as coherence, LQU and  UIN, we plot these measures in Fig (1b \& c) as a function of Unruh temperature. As Unruh temperature $T_U$ increases, the quantum correlation measures decrease initially and reach  a dark point (zero). While increasing the temperature $T_U$ further, both LQU and coherence start to grow from the dark point and then correlations increase with the Unruh temperature $T_U$ unconventionally. It is quite obvious that both the measures are non-monotonic  functions of $T_U$.  In general, if we immerse any static two-qubit in a thermal bath, the quantumness of the system degrades with the temperature of the thermal bath. But here, on the contrary, we observe the revival of quantum correlations and coherence due to the acceleration of the qubits. In addition, the dark point is shifted towards higher temperatures with the increase of $\Delta_0$. Similar to entanglement, UIN decreases from the maximum values with the increase of temperature and saturates at a nonzero constant value.  A similar observation is noticed using trace distance correlation measure \cite{Huang1}.

 We observe the nonlocal features between the detectors even in the absence of entanglement which are captured through the $C_{l_1}$- norm of coherence and skew information based measures. The entanglement fails  to completely quantify the nonlocality of the UdW detector for $\Delta_0=1$. On the other hand, the correlation measures beyond entanglement are encapsulated in UdW detector even in the absence of entanglement. 

Next, we analyze the role of energy spacing $\omega$ on the quantum correlation measures. For this purpose, we plot them as a function of  Unruh temperature $T_U$ with specific energy spacing of detectors $\omega=1,3,5$.  It is obvious that entanglement always degrades monotonously with increasing Unruh temperature, i.e., no revival of entanglement can happen for any initial state preparation. This clearly demonstrates the distinction between the entanglement and coherence.  In Fig. (\ref{fig1})a, we observe that the entanglement vanishes at low temperatures $T_U=0.5$ for $\Delta_0=0$. For the same initial state $(\Delta_0=0)$, comparing Fig. (\ref{fig1})a with Fig. (\ref{fig2})a, we observe that the entanglement is induced in the higher temperature regime while increasing the values of energy spacing of detectors from $\omega=1$ to $\omega=3$. If we  increase the value of $\omega$ further, one  observes similar monotonic decreasing behavior of the correlation measures while the entanglement survives for sufficiently higher temperatures  compared to lower values of $\omega$. On the other hand, the other correlation measures also increase with the increase of energy spacing. In Fig. (\ref{fig3}), we have plotted the correlation measures as a function of $T_U$ for another  initial state such as $\Delta_0=0.5$. Here again, the entanglement decreases monotonously while  one sees the revival of LQU and coherence measures.


\begin{figure}[t]
\begin{center}
\includegraphics[width=.45\textwidth]{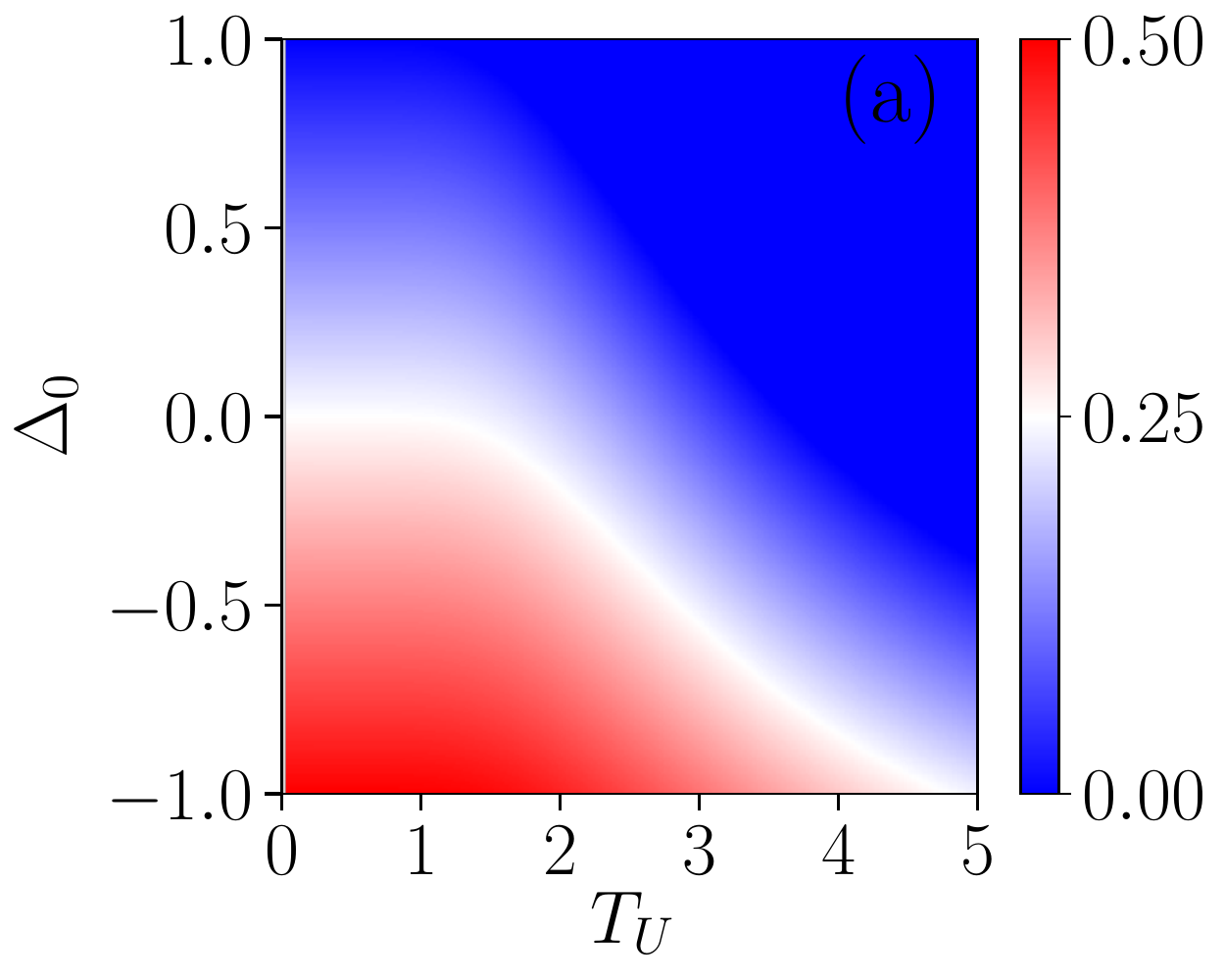}
\includegraphics[width=.45\textwidth]{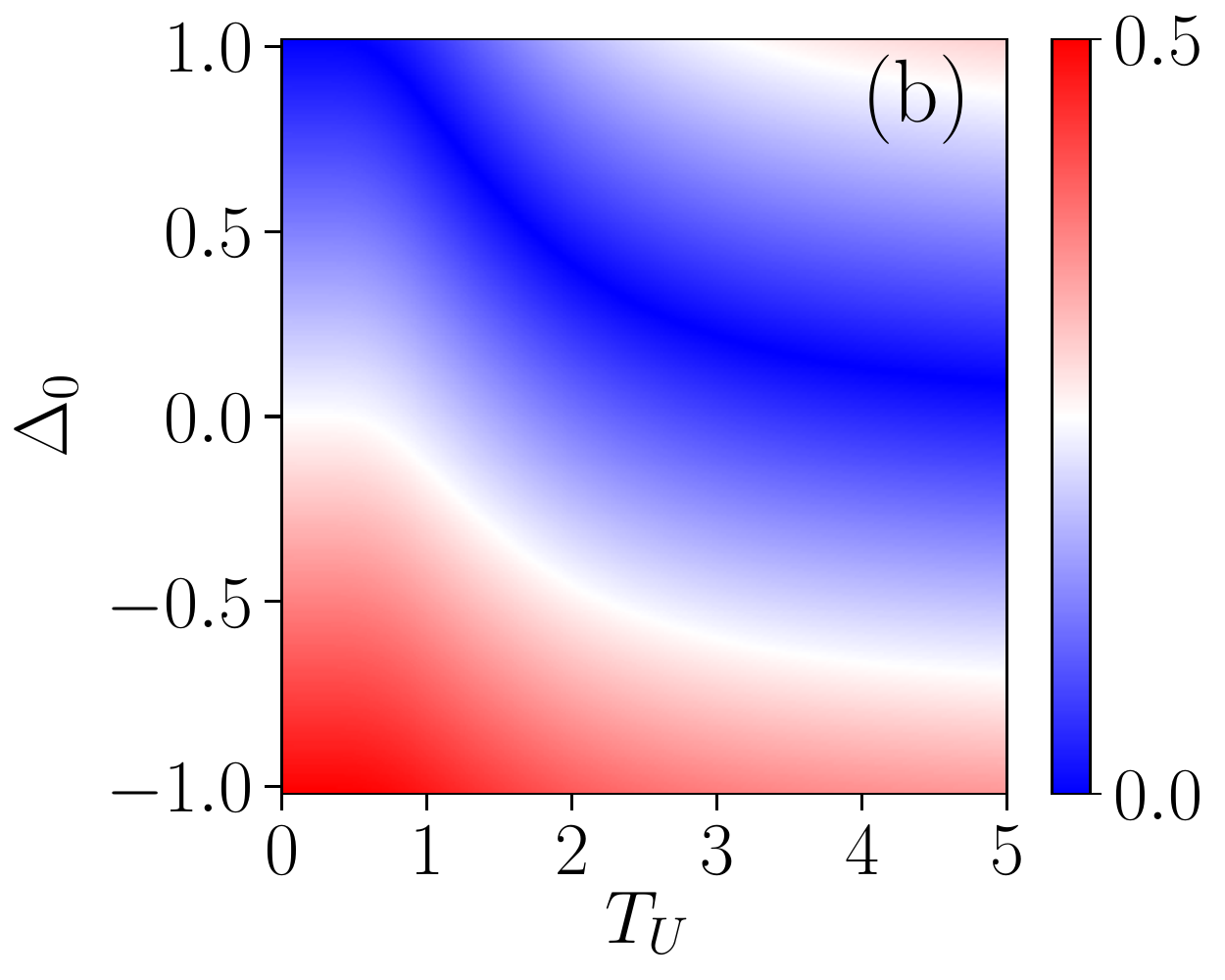}
\includegraphics[width=.45\textwidth]{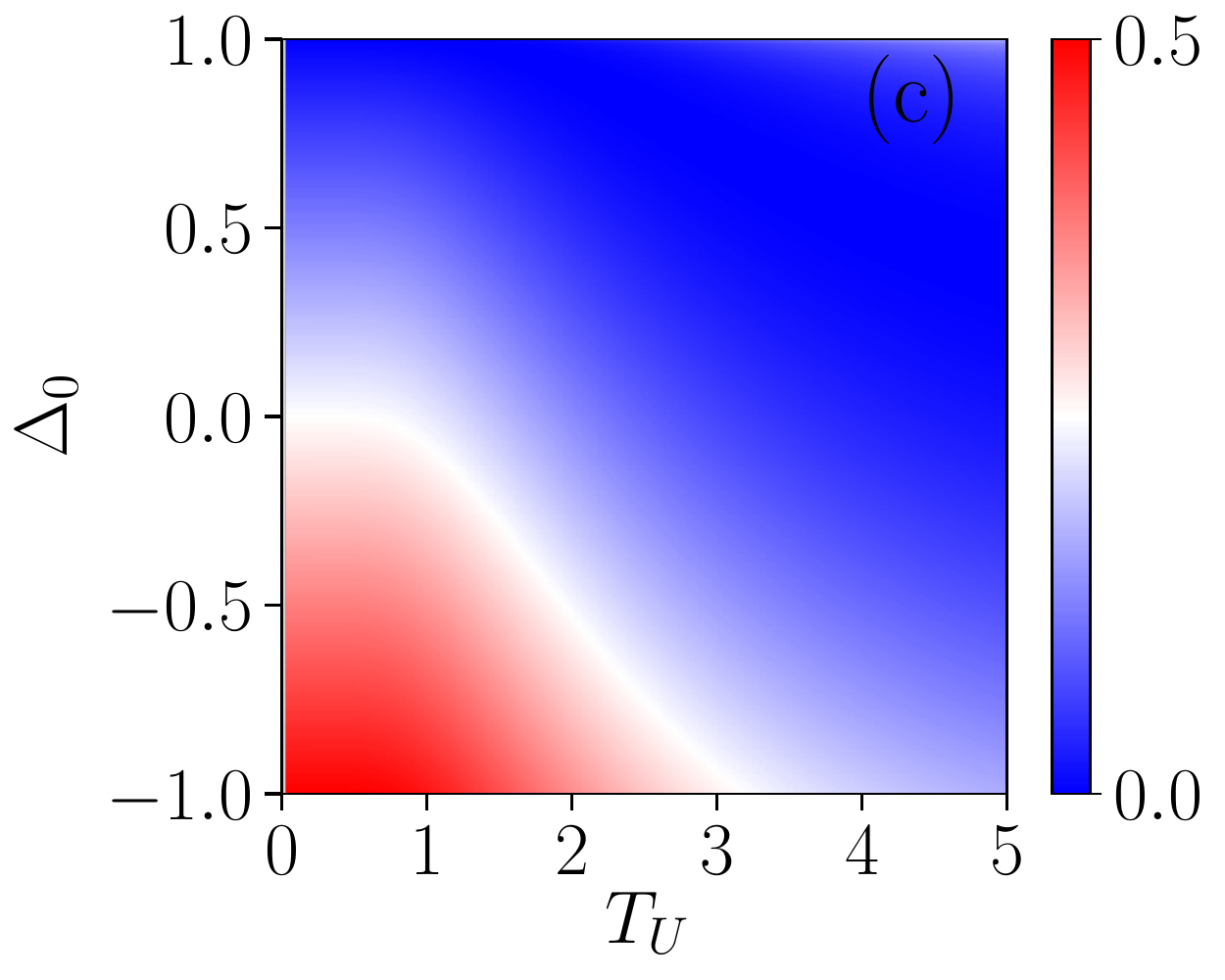}
\includegraphics[width=.45\textwidth]{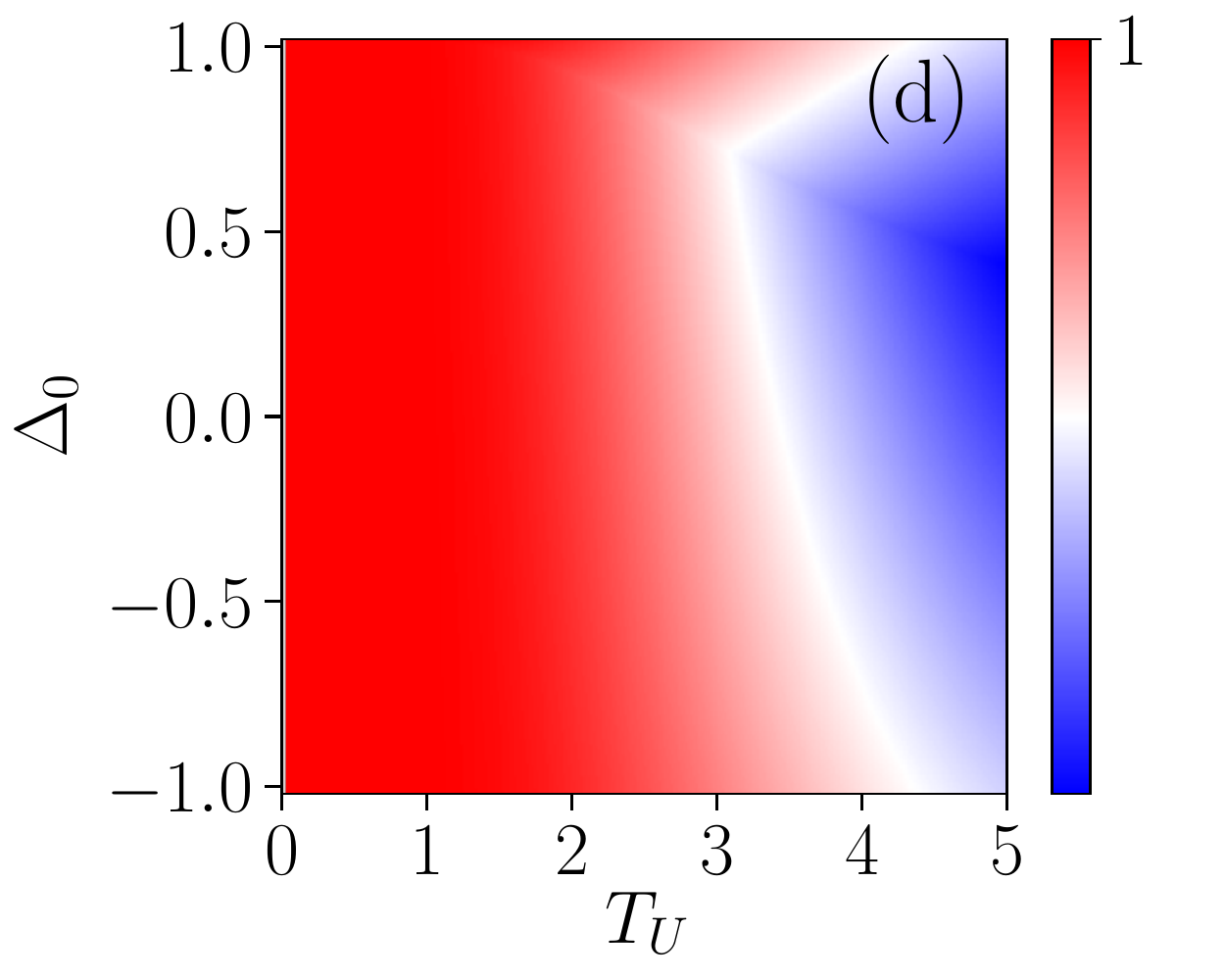}
\caption{The Density of quantum correlation measures (a) Coherence, (b) LQU, (c) Entanglement and (d) UIN of UDW detector as a function of $T_U$ and $\Delta_0$ for   $\omega=3$}
\label{fig5}
\end{center}
\end{figure}

Now, we illustrate our results with  an example. If we choose the initial state of two-detector in a product form 
\begin{align}
\rho_{\text{in}}=\rho_a(0)\otimes\rho_b(0),      \label{eq19}
\end{align}
 correlations of the initial state become zero. 
 The state of each detector can be written in Bloch form as
\begin{align}
\rho_{a}(0)=\frac{1}{2}\left(I+n \cdot \sigma \right),~~~\rho_{b}(0)=\frac{1}{2}\left(I+m \cdot \sigma \right)   \label{eq20}
\end{align}
where $\mathbf{n}$ and $\mathbf{m}$ are two unit Bloch vectors. Without loss of generality, taking $\mathbf{n}=(0,0,1)$ and $\mathbf{m}=(0,\sin\theta,\cos\theta)$, we have $\Delta_0=\mathbf{n}\cdot\mathbf{m}=\cos\theta$ where $\theta\in[0,\pi]$ is an angle between two vectors giving $\Delta_0\in[-1,1]$. 

In Fig. (\ref{fig5}), we have plotted the density of   quantum correlation measures as a function of $T_U$ and $\Delta_0$ with $\omega=3$. Again, it is obvious  that the entanglement is generated during time evolution of the detectors for $\Delta_0=(-1,0)$ and there is no entanglement induced in the range $\Delta_0=(0,1)$.  Here again, we notice that the entanglement  vanishes at low temperatures and is unable to   capture the complete manifestation of nonlocal attributes of UD detectors. On the other hand, the quantum correlation measures  quantify more nonlocal aspects of UD detectors compared to entanglement. It can be noticed that  the generation of entanglement and quantum correlations in the initial product state arises due to the acceleration of the system.

\section{Conclusions}
\label{conc}
In this paper, we have studied the  quantum coherence and correlations  of two accelerating Unruh-
DdeWitt detectors coupled to a scalar background in 3 + 1 Minkowski spacetime. We have employed the entanglement, local quantum uncertainty, uncertainty-induced nonlocality and $l_1$-norm of coherence as the quantumness quantifiers. Results of the paper indicate the revival of  LQU and coherence  while entanglement does not. We have also shown that the energy spacing of detectors induces the nonlocality between the detectors.  The distinction between the static and accelerating qubit systems is observed in terms of generation of quantumness in a product state and revival mechanism of quantum correlations.

Further, the results of our manuscript can have wider ramifications in understanding  the relativistic quantum information processing from the perspective of quantum correlation measures.
\noindent

\section*{Acknowledgment}

Authors are indebted to the referees for their critical comments to improve the contents of the manuscript. SB and RR wish to  acknowledge the financial support received from the Council of Scientific and Industrial Research (CSIR), Government of India under Grant No. 03(1456)/19/EMR-II.

\end{document}